\documentclass[12pt]{elsarticle}
\usepackage{amssymb}
\usepackage{amsmath}
\usepackage{graphics}
\usepackage{latexsym}
\usepackage{array}
\usepackage[activeacute,english]{babel}
\usepackage[latin1]{inputenc}
\usepackage{float} % Required for tables and figures in the multi-column environment - they need to be placed in specific locations with the [H] (e.g. \begin{table}[H])
\usepackage{color}
\usepackage{latexsym}
\usepackage[colorlinks=true]{hyperref}

\newcommand{\bn}{\begin{eqnarray}}
\newcommand{\en}{\end{eqnarray}}
\newcommand{\bq}{\begin{equation}}
\newcommand{\eq}{\end{equation}}

\newcommand{\bc}{\begin{center}}
\newcommand{\ec}{\end{center}}

\begin{document}

\begin{frontmatter}

\title{Long-range correlations and trends in Colombian seismic time series}

\author[desy,unc]{L. A. Mart\'in-Montoya}
\ead{ligia.andrea.martin.montoya@desy.de}

\author[usp,unc]{N. M. Aranda-Camacho}
\ead{nataly@iag.usp.br}

\author[unc,cif]{C. J. Quimbay}
\ead{cjquimbayh@unal.edu.co}

\address[desy]{Deutsches Elektronen-Synchrotron, Notkestr. 85, 22607
Hamburg, Germany}
\address[usp]{University of S\~ao Paulo, Rua do Mat\~ao 1226,
05508-090, S\~ao Paulo, SP, Brazil}
\address[unc]{Departamento de F\'\i sica, Universidad Nacional de Colombia,
Bogot\'a, D. C., Colombia}
\address[cif]{Associate researcher of CIF, Bogot\'a, Colombia}
\begin{abstract}
We study long-range correlations and trends in time series
extracted from the data of seismic events occurred from 1973 to
2011 in a rectangular region that contains mainly all the
continental part of Colombia. The long-range correlations are
detected by the calculation of the Hurst exponents for the time
series of interevent intervals, separation distances, depth
differences and magnitude differences. By using a modification of
the classical $R/S$ method that has been developed to detect
short-range correlations in time series, we find the existence of
persistence for all the time series considered except for
magnitude differences. We find also, by using the $DFA$ until the
third order, that the studied time series are not influenced by
trends. Additionally, an analysis of the Hurst exponent as a
function of the number of events in the time and the maximum
window size is presented.

\end{abstract}

\end{frontmatter}

%%PACS:03.65.Ge, 03.65.Pm
\newpage
\section{Introduction}\label{sec:01}

Among the properties that time series of different kind of phenomena
exhibit, one of the most interesting is the long-range correlation
\cite{Stanley1971}. This correlation, also known as long memory or
long-range persistence, means that the autocovariance function
decays exponentially, by a spectral density that tends to infinity
\cite{Mantegna2000,Grau2000}. However, at the critical point, the
exponential decay turns into a power-law decay
\cite{Mantegna2000,Varotsos2002}. Long-range power-law
correlations can be observed in a very wide range of systems, as
for instance in smoke-particle aggregates \cite{Forrest1979},
nucleotide sequences \cite{Peng1992}, earthquake processes
\cite{Hirabayashi1992}, mosaic structure of DNA \cite{Peng1994},
literary texts \cite{Ebeling1994}, stride interval of human gait
\cite{Hausdorff1995}, cardiac interbeat intervals \cite{Peng1995},
stock index variations \cite{Liu1997}, sedimentation
\cite{Segre1997}, variations of daily maximum temperatures
\cite{Koscielny1998}, neuron activity \cite{Blesic1999}, stratus
cloud liquid water \cite{Ivanova1999}, stock returns
\cite{Grau2000}, electric signals \cite{Varotsos2002}, seismic
sequences \cite{Telesca2004-1,Telesca2004-2}, and relativistic
nuclear colisions \cite{Gavin2009}.

Some geophysical processes are characterized by self-similarity
correlations because the dynamics are based on the interaction
of many components over a wide range of time or space scales \cite{Ashkenazy}.
An example of complex systems in geophysics are earthquakes, characterized by
power-law distributions of spatial, temporal and energy parameters \cite{Telesca2008}.
The complex phenomenology exhibited by earthquakes is due to the
deformation and sudden rupture of parts of the earth's crust because
of the external forces acting from plate tectonic motions \cite{Michas}.

Long-range power-law correlations are traditionally measured by a
scaling parameter or fractal dimension $(D)$. If the time series
is self-similar and self-affine, the parameter $D$ is related
to the Hurst exponent ($H$) through the expression $D=2-H$
\cite{Feder1988,Chen2008}. Thus, the Hurst exponent is a measure
of the long-range correlation in time series data and allows to
distinguish the persistence (correlation), anti-persistence
(anti-correlation) or randomness of the data
\cite{Mandelbrot1968}. The original estimation of the Hurst
exponent was first performed in hydrology by Harold Edwin Hurst in
1951 \cite{Hurst1951}, by introducing an empirical relationship
called the Rescaled-Range $(R/S)$. Posteriorly, this relationship
became the start point to establish the Classical $R/S$ ($CR/S$)
method developed by Mandelbrot and Wallis into the context of the
fractal geometry \cite{Mandelbrot1968, Mandelbrot1969-1,
Mandelbrot1969-2}.

Although the $CR/S$ is one of the most popular methods to
calculate the Hurst exponent, it has shown some serious
limitations to study long-range correlation when the time series
is not large enough \cite{Sanchez2008, Trinidad2012}. However, a
possible solution of this problem was proposed by W. Lo, which we
call the Modified R/S $(MR/S)$ method \cite{Lo}. Another of the
most popular methods to calculate de Hurst exponent is the
Detendred Fluctuation Analysis $(DFA)$ proposed by Peng {\it et
al.} \cite{Peng1994}. One of the main advantages of the $DFA$ is
that it can remove trends present in some real time series
\cite{Kantelhardt2001,Hu2001,Eichner2003}. Additionally, this
method can be generalized for the multifractal characterization of
nonstationary time series \cite{Kantelhardt2002}.

The study of long-range power-law correlations in seismic
sequences has been focussed on the calculation of the Hurst
exponent for time series of interevent intervals by using the
$CR/S$ \cite{Chen2008} and the $DFA$
\cite{Telesca2004-1,Telesca2004-2}. The DFA was also used to
calculate the Hurst exponent associated to two-dimensional
sequences defined in terms of the time series of interevent
intervals and separation distances \cite{Telesca2007}.
Furthermore, other time series like temporal and spatial
variations were studied using the $CR/S$ method
\cite{Alvarez2012}. Persistence in temporal sequences for
earthquake activity has been recognized and reported in a number
of studies \cite{Telesca2004-1,Telesca2004-2,Chen2008,Alvarez2012}
and also for the temporal-spatial sequences \cite{Telesca2007}.
This result is consistent with the fractal structures in time,
space, and magnitude dimensions observed in the seismicity of
earthquakes \cite{Hirata1989,Guo1995,Guo1997}. However, most
recently, an anti-correlated behaviour $(0<H<0.5)$ of the
earthquake magnitude time series was observed from the calculation
of the Hurst exponent using the $DFA$ \cite{Varotsos2012}

Until now the calculation of the Hurst exponent in seismic time
series has been focussed in temporal, temporal-spatial and
magnitude sequences
\cite{Telesca2004-1,Telesca2004-2,Chen2008,Telesca2007,Alvarez2012,Varotsos2012}.
In this work we extend the detection of long-range correlations
for the case of new time series that can be extracted from the
seismic event data. Additionally to the time series of interevent
intervals
\cite{Telesca2004-1,Telesca2004-2,Chen2008,Alvarez2012,Jimenez2011},
we consider also the time series associated to the following
parameters defined between two successive seismic events:
separation distance \cite{Hirata1989,Telesca2007}; depth
difference; magnitude difference. We think that if we study
simultaneously the existence of long-range correlations for these
four seismic time series, then we have a most complete vision of
the fractal structure associated to the seismic events
\cite{Hirata1989}. Because the interaction between a previous
event and the event caused by it relaxes with the time, it is
possible to observe simultaneously a self-similar structure in
time, space and magnitude distributions of the seismic event
epicentres \cite{Hirata1989,Guo1995,Guo1997}.

The main goal of this work is to detect whether long-range
correlations and trends exist in time series extracted from the
data of seismic events that occurred since 1973 until 2011 in a
rectangular region that contains mainly all the continental part of
Colombia. During this lapse of 39 years, the region under study,
which contains also parts of Brazil, Ecuador, Panama, Peru and
Venezuela, presented a total of 3932 seismic events with
magnitudes higher than 4 $M_W$ according to the U.S. Geological
Survey data base \cite{usgs}. To detect the existence of
long-range correlations in the time series of interevent
intervals, separation distances, depth differences and magnitude
differences, we calculate the Hurst exponents associated to these
time series by using the $MR/S$ method
\cite{Lo}. To appreciate the advantage of
using the $MR/S$ method, we first calculate the Hurst exponents
for all the time series by using the $CR/S$ method. With the
purpose to eliminate trends in the seismic time series considered
here, we calculate the Hurst exponents using the $DFA$ until the
third order \cite{Kantelhardt2001,Hu2001,Kantelhardt2002}.

The reason why we use the $MR/S$ \cite{Lo} is because the $CR/S$
might lead to an overestimation of the Hurst exponent, specially
if seismic time series are constructed with less than 5000 data
points as happened in this work. On the other hand, a study
previously performed using data of Southern M\'exico
\cite{Alvarez2012} has detected the existence of cycles in the
interevent interval sequence and this fact has been considered as
a sign of trends in the associated seismic time series. For this
reason, in this work, we use also the $DFA$ with the purpose to
find polynomial trends on the data. We study also the evolution on
time of the correlation between the Hurst exponent and the
magnitude for the interevent interval in the seismological active
area of the subduction zone of the Nazca and the South American
plates in Colombia. The goal of this part of our study is to
analyse if there is a statistical correlation between the value of
the Hurst exponent and the magnitude of the seismic event, as it
was found for the time series of interevent intervals in Southern
M\'exico \cite{Alvarez2012}. For accurateness, a study on the
maximum window size in the calculation of the Hurst exponent is
also performed, focusing the interest on the Hurst exponent as a
function of the number of events in the time and the maximum
window size.

%%%%%%%%%%%%%%%%%%%%%%%%%%%%%%%%%%%%%%%%%%%%%%%%%%%%%%%%%%%%%%%%%%
%%%%%%%%%%%%%%%%%%%%%%%%%%%%%%%%%%%%%%%%%%%%%%%%%%%%%%%%%%%%%%%%%%
\section{Methods to calculate the Hurst exponent}
%%%%%%%%%%%%%%%%%%%%%%%%%%%%%%%%%%%%%%%%%%%%%%%%%%%%%%%%%%%%%%%%%%
%%%%%%%%%%%%%%%%%%%%%%%%%%%%%%%%%%%%%%%%%%%%%%%%%%%%%%%%%%%%%%%%%%

%%%%%%%%%%%%%%%%%%%%%%%%%%%%%%%%%%%%%%%%%%%%%%%%%%%%%%%%%%%%%%%%%%
%%%%%%%%%%%%%%%%%%%%%%%%%%%%%%%%%%%%%%%%%%%%%%%%%%%%%%%%%%%%%%%%%%
\subsection{Classical Rescaled Range (CR/S)}

The $CR/S$ method allows to study the long-range dependence for
non-stationary time series by means of the calculation of the
Hurst exponent $H$ \cite{Mandelbrot1968, Mandelbrot1969-1,
Mandelbrot1969-2}. Specifically, a value $0<H<0.5 $ corresponds to
anti-correlated data (anti-persistence behaviour), a value $0.5<H<1 $
corresponds to correlated data (persistence behaviour) and the
value $H=0.5$ corresponds to random data (uncorrelated behaviour)
\cite{Mandelbrot1968}. The definition of $H$ can be extended to
values larger than $1$ \cite{Jimenez2011}. In this way, the case
$H=1.5$ corresponds to Brownian motion, the case $H=2$ correspond
to brown noise and the case $H>2$ corresponds to black noise
\cite{Jimenez2011}.

In the $CR/S$ method the time series under consideration
{\footnotesize$X:\{x_i\}$} is composed by {\footnotesize$N$}
values. The full time series is divided into windows of size
{\footnotesize$M$}. The number of windows is defined by
{\footnotesize$s \equiv N/M$} and therefore there are
{\footnotesize$s$} windows of data {\footnotesize$Y_j$}, with
{\footnotesize$j=1,2,...,s$}.
\\
\newline
Defining the vector {\footnotesize$k=(j-1)M+1, (j-1)M+2,
(j-1)M+3,...,(j-1)M+M$}, the average over each window is
calculated as {\footnotesize
\begin{equation}
 \bar{y}_{j}=\frac{1}{M}\sum_{k}x_k.
\end{equation}
}The profile or sequence of partial summations
{\footnotesize$Z_j:\{z_n\} $}, with {\footnotesize$n=1,2,..., M$},
is defined as the cumulative summation minus the average of the
corresponding window
 {\footnotesize
 \begin{equation}
  z_n=\sum_{k}^{n}\{x_k-\bar y_j\}.
 \end{equation}
}The range $R_j$ of the window is defined as the maximum minus the
minimum data point of the profile {\footnotesize
 \begin{equation}
  R_j\equiv max\{Z_j\}-min\{Z_j\}.
 \end{equation}
}The standard deviation of each window $\sigma_j$ is given as
{\footnotesize
\begin{equation}
 \sigma_j=\left[ \frac{1}{M} \sum_{k} \left(x_k-\bar y_j\right)^2\right]^{1/2}.
\end{equation}
}The rescaled range is described by the quantity
{\footnotesize$(R/S)_M$}, which is defined as {\footnotesize
\begin{equation}
 \left(R/S\right)_M\equiv mean(R_j/ \sigma_j).
\end{equation}
}For the case in which a stochastic process associated to the data
sequence under study is rescaled over a certain domain
{\footnotesize$M \in \{ M_{min},M_{max} \}$}, the
{\footnotesize$R/S$} statistics follows the power law
{\footnotesize
\begin{equation}
 (R/S)_M=a M^H.
\end{equation}
}Herein, {\footnotesize$a$} is a constant and {\footnotesize$H$}
is the Hurst exponent which represents a fractal measure of the
long-range correlations in the analysed data.

%%%%%%%%%%%%%%%%%%%%%%%%%%%%%%%%%%%%%%%%%%%%%%%%%%%%%%%%%%%%%%%%%%
%%%%%%%%%%%%%%%%%%%%%%%%%%%%%%%%%%%%%%%%%%%%%%%%%%%%%%%%%%%%%%%%%%

\subsection{Modified Rescaled Range (MR/S)}

When dealing with short time series (of less than 5000 values),
the $CR/S$ method looses precision \cite{Sanchez2008}. In order to
study short time series the R/S statistics should be modified so
that it accounts for short memory processes \cite{Lo}. When the
time series under consideration is subject of short-range
dependence, the variance should also include the auto-variances.
This is accomplished when the standard deviation term described in
equation (4) is replaced by the term:

\begin{equation}
\begin{split}
 \sigma_j(q)=\sqrt{\sigma_j^2+2\sum_{k=1}^{q}\omega_k(q)\gamma_k},
 \\
 \omega_k(q)=1-\frac{k}{q+1}
 \end{split},
\end{equation}

where $\sigma_j^2$ is the sample variance and $\gamma_k$ the
sample autocovariance. The parameter $q$  represents the lags in
the weighted autocovariances.

%%%%%%%%%%%%%%%%%%%%%%%%%%%%%%%%%%%%%%%%%%%%%%%%%%%%%%%%%%%%%%%%%%
%%%%%%%%%%%%%%%%%%%%%%%%%%%%%%%%%%%%%%%%%%%%%%%%%%%%%%%%%%%%%%%%%%
\subsection{Detrended Fluctuation Analysis (DFA)}

An important tool to leave out trends in time series is the $DFA$
\cite{Kantelhardt2001,Hu2001,Eichner2003}. Trends can affect the
time series and make them seem to be persistent when there is not
a real correlation between the data points
\cite{Kantelhardt2001,Hu2001,Eichner2003}.
In order to find the correct scaling behaviour of the
series, such trends have to be well distinguished. It is not easy
to identify the origin of trends, so very often we do not know
the reasons for underlying trends in collected data and we do not know the
scales of underlying trends. {\it DFA} is a method for determining
the scaling behaviour of data in the presence of possible
trends without knowing their origin and shape \cite{Telesca2008}.
In the following, the
method to calculate the Hurst exponent via $DFA$ is explained.

As the first step of the $DFA$, it is necessary to
calculate the profile or cumulative summations minus the average
of all the data {\footnotesize$\bar x_i$}
\\
 {\footnotesize
\begin{equation}
 z_k= \sum _{i=1}^{k}x_i- \bar x_i,
\end{equation}}
where {\footnotesize$k$} runs over each data point
{\footnotesize$k=1,2,3,...,N$}.

As a second step, the profile is divided into {\footnotesize$s$}
windows of size {\footnotesize $M$} and a fit of order
{\footnotesize$l$} of the window is subtracted from each data
point. Choosing the order to be {\footnotesize$l=1$}, each new
data point {\footnotesize$y_i$} is described as {\footnotesize
\begin{equation}
 y_k=z_k-\{a_j\cdot k-b_j\},
\end{equation}
}where {\footnotesize$a_j$} and {\footnotesize$b_j$} are
respectively the slope and the intercept of the linear fit done
over the window {\footnotesize$j$}. For this case, the subscript
{\footnotesize$j$} runs over the windows
{\footnotesize$j=1,2,3,...,s$}. The fluctuation is calculated as
{\footnotesize
\begin{equation}
 F_M=\sqrt{\left[\frac{1}{M}\sum_{k} y_k^2\right]}.
\end{equation}
}Varying the size of the window {\footnotesize $M$} and
consequently the number of windows {\footnotesize $s$}, the Hurst
exponent {\footnotesize $H$} corresponds to the slope in the
log-log plot of {\footnotesize $F_M$}. If a stochastic process is
associated to the data sequence under study, it will follow the power
law
\\
{\footnotesize
\begin{equation}
 F_M=aM^H.
\end{equation}
}

Different hierarchies of methods are employed for the fluctuation
analysis. Those hierarchies differ from each other in how the
fluctuations are measured and how trends are eliminated. By
definition {\footnotesize$DFA_n$} eliminates trends of order
{\footnotesize$n-1$} in the original time series and
{\footnotesize$n$} in the profile
\cite{Kantelhardt2001,Hu2001,Eichner2003}. For instance,
{\footnotesize$DFA_0$} corresponds to the simplest type of
fluctuation analysis and in this case the trends are not
eliminated. In the first order detrended fluctuation analysis
{\footnotesize$DFA_1$}, the variance of the profile in each window
represents the square of the fluctuations and then the best linear
fit is determinated \cite{Eichner2003}.
\\

%%%%%%%%%%%%%%%%%%%%%%%%%%%%%%%%%%%%%%%%%%%%%%%%%%%%%%%%%%%%%%%%%%
%%%%%%%%%%%%%%%%%%%%%%%%%%%%%%%%%%%%%%%%%%%%%%%%%%%%%%%%%%%%%%%%%%
\section{Data acquisition}
%%%%%%%%%%%%%%%%%%%%%%%%%%%%%%%%%%%%%%%%%%%%%%%%%%%%%%%%%%%%%%%%%%
%%%%%%%%%%%%%%%%%%%%%%%%%%%%%%%%%%%%%%%%%%%%%%%%%%%%%%%%%%%%%%%%%%

\begin{figure}[H]
\begin{center}
% For example, with the graphicx package use e.g
\includegraphics[width=0.45\textwidth]{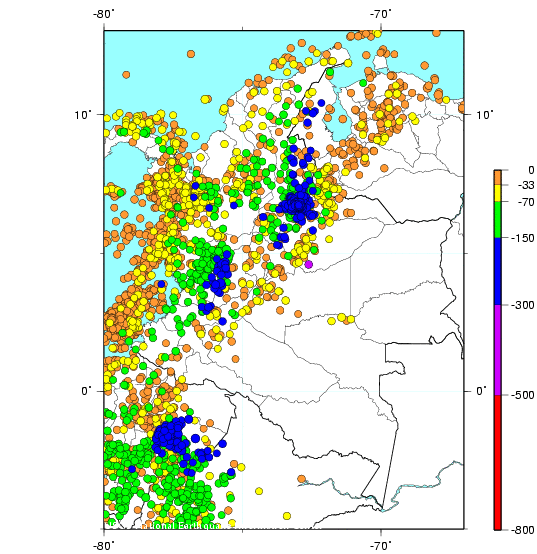}
\caption{Selected region for analysis of seismic data. Events
occurring from 1973 to 2011 are shown as small circles with
different colors accordingly with the depth of the event
\cite{usgs}.}
\label{fig2} % Give a unique label
\end{center}
\end{figure}

We consider in this work the time series associated to the
following parameters defined between two successive seismic
events: {\it (i)} interevent interval
\cite{Telesca2004-1,Telesca2004-2,Chen2008,Alvarez2012,Jimenez2011};
{\it (ii)} separation distance \cite{Hirata1989,Telesca2007}; {\it
(iii)} depth difference; {\it (iv)} magnitude difference. We note
that the seismic time series of depth differences and magnitude
differences are new for this kind of analysis. The analysis
for time series of separation is new because we consider the time series of
separation distances as independent respect to the time series of
interevent intervals different from reference \cite{Telesca2007}.

We analyzed the above mentioned time series because we are interested
in knowing the scaling behavior of the most important characteristics of earthquakes.
We think that knowing the behavior between, for example, the depth or
the epicenter distance, or the difference in magnitude of two
consecutive events in an area, would help to better describe the
fractal behavior of the earthquakes.

The time series of separation distances is extracted from the
seismic event data using the following definition for the angular
distance $r$ in degrees between two events
\cite{Hirata1989,Telesca2007}
\begin{equation}
r=\cos^{-1}[\cos(\theta_1)\cos(\theta_2)+\sin(\theta_1)\sin(\theta_2)
\cos(\phi_1-\phi_2)],
\end{equation}
where $(\theta_1,\theta_2)$ and $(\phi_1,\phi_2)$ are respectively
the colatitudes and the longitudes of the two events
\cite{Hirata1989,Telesca2007}. If we multiply the angular
distance $r$ by 111 km, then we obtain the separation distance
between two successive seismic events \cite{Telesca2007}.

The time series of interevent intervals, separation distances,
depth differences and magnitude differences are constructed from
the seismic event data that we have obtained from the free online
database of the United States Geological Survey \cite{usgs}. A
rectangular region that contains mainly all the continental area
of Colombia was chosen to analyse the seismic data occurred from
1973 to 2011. As it is possible to observe in Figure 1, this
region also contains parts of Brazil, Ecuador, Panama, Peru and
Venezuela. Seismic events with magnitudes higher than 4 $M_W$ were
selected in such a way that the total number of events initially
considered was 3932. Using these data, Figure 2 presents the plot
of the logarithm of the frequency as a function of the magnitude
of the seismic event. The blue line in Figure 2 underlines the
events with magnitudes from 4.8 to 6 $M_W$ that correspond to the
ones that satisfy the Gutenberg-Richter law. Those events define
the set that is used to perform the Hurst exponential analysis
that will be presented below. It is important to mention that this
methodology based on the Gutenberg-Richter law has been widely
used in this kind of analysis
\cite{Telesca2004-1,Telesca2004-2,Telesca2007,Chen2008,Alvarez2012}.

\begin{figure}[H]
\begin{center}
% For example, with the graphicx package use e.g
\includegraphics[width=0.45\textwidth]{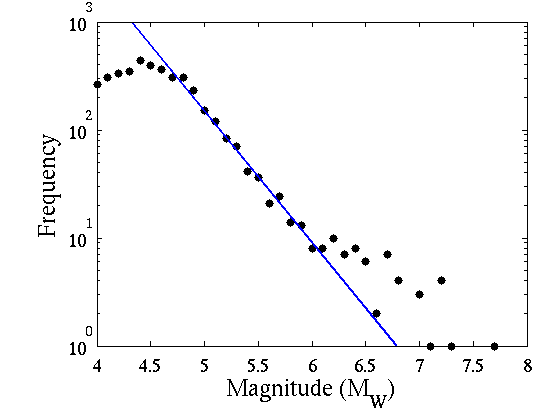}
\caption{Logarithm of frequency vs. magnitude for the seismic
events occurred since 1971 until 2011 in the selected region. The
blue line shows the events that satisfy the Gutenberg-Richter law.
Only events from 4.8 to 6 $M_W$ were used to perform the Hurst
exponential analysis.}
\label{fig2} % Give a unique label
\end{center}
\end{figure}

%%%%%%%%%%%%%%%%%%%%%%%%%%%%%%%%%%%%%%%%%%%%%%%%%%%%%%%%%%%%%%%%%%
%%%%%%%%%%%%%%%%%%%%%%%%%%%%%%%%%%%%%%%%%%%%%%%%%%%%%%%%%%%%%%%%%%
\section{Persistence and trends in Colombian seismic data}
%%%%%%%%%%%%%%%%%%%%%%%%%%%%%%%%%%%%%%%%%%%%%%%%%%%%%%%%%%%%%%%%%%
%%%%%%%%%%%%%%%%%%%%%%%%%%%%%%%%%%%%%%%%%%%%%%%%%%%%%%%%%%%%%%%%%%

Long-range correlations for temporal and temporal-spatial
sequences have been detected previously by the calculation of
Hurst exponent for different seismic active areas of the world, as
for instance of Italy \cite{Telesca2004-1,Telesca2004-2}, USA
\cite{Telesca2007,Varotsos2012,Jimenez2011}, Taiwan
\cite{Chen2008}, M\'exico \cite{Alvarez2012}, and Greece
\cite{Burton2006}. In this section we present the results of the
calculation of the Hurst exponent for the Colombian seismic time
series of interevent intervals, separation distances, depth
differences and magnitude differences. Firstly, we present the
Hurst exponents that were calculated using both the $CR/S$ and
$MR/S$ methods. By using the $MR/S$ method, we obtain Hurst
exponents more precise, correcting for the effect of insufficiency of
data (less than 5000 data points). Secondly, we present the Hurst
exponents calculated using the $DFA$ until the third order. In
this way, we can remove the trends which are present in the
seismic time series under study.

%%%%%%%%%%%%%%%%%%%%%%%%%%%%%%%%%%%%%%%%%%%%%%%%%%%%%%%%%%%%%%%%%%
%%%%%%%%%%%%%%%%%%%%%%%%%%%%%%%%%%%%%%%%%%%%%%%%%%%%%%%%%%%%%%%%%%
\subsection{Hurst exponents and scarcity effects}

A correct degree of long-range correlation, represented by a
precise calculation of the Hurst exponent, in time series composed
by a relative small quantity of data can not obtained by using the
$CR/S$ method.

We have tested 1000 random time series of 100 events for both the
$CR/S$ and the $MR/S$. Results in figure 3 show that for random
time series there is an overestimation of the Hurst exponent
calculated with the $CR/S$. This method yields $H_C$ of around
0.6, while with the $MR/S$ this overestimation is corrected giving
$H_M$ of around 0.55.

\begin{figure}[H]
\begin{center}
% For example, with the graphicx package use e.g
\includegraphics[width=0.45\textwidth]{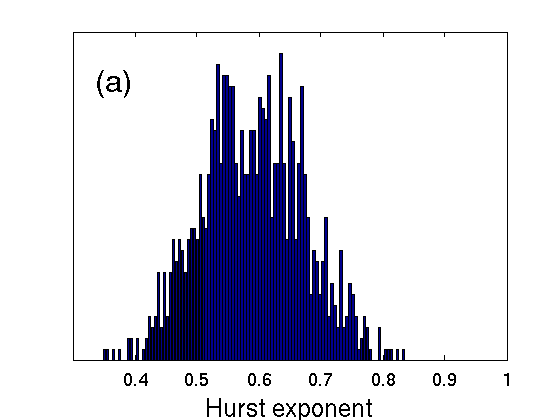}
\includegraphics[width=0.45\textwidth]{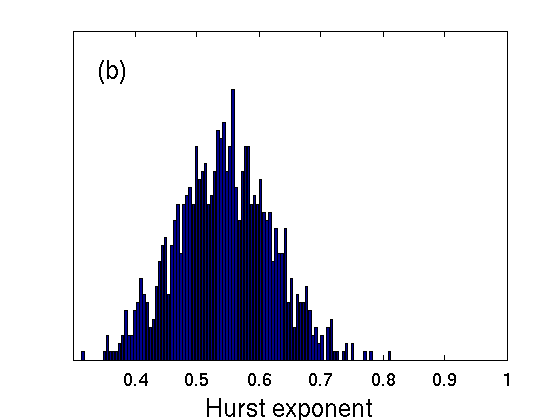}
\caption{Distribution of the Hurst exponent for 1000 random time series of length 100.
 {\bf(a)} Hurst exponent calculated
with the $CR/S$ method. {\bf(b)} Hurst exponent calculated using
the $MR/S$ method. }
\label{fig3} % Give a unique label
\end{center}
\end{figure}

The fact that the $CR/S$ method leads to an overestimation of the
Hurst exponent for the considered seismic time series means
that a fake degree of long-range correlation is originated by an
insufficient number of data. This inaccuracy of the Hurst exponent
calculated using the $CR/S$ method for less than 5000 data points
was proved for capital markets \cite{Sanchez2008}. We find that if
the four seismic time series are analysed using the $MR/S$ method,
then the Hurst exponents ($H_{M}$) obtained with this modified
method are smaller than the ones calculated by the $CR/S$ method
($H_{C}$). We show in the Table 1 the differences between the
Hurst exponents calculated using both methods, for the time series
of interevent intervals, separation distances, depth differences
and magnitude differences.

For this calculations parameters of $M_{max}$ were chosen by
visual inspection of the $log(R/S)-log(M)$ plots so that they show
a linear dependence in each of the cases, while parameters of
$M_{min}$ were chosen equal to 5 in all calculations. By doing
this, we assure that the data satisfy the power law in the chosen
range of $M_{nim}-M_{max}$.

\begin{center}
\begin{tabular}{| l l l l l l|}

\hline
\scriptsize{Time series} &   \scriptsize{$H_{C}$} & \scriptsize{$H_{M}$}  &   \scriptsize{$H_{DFA_{1}}$} & \scriptsize{$H_{DFA_{2}}$}& \scriptsize{$H_{DFA_{3}}$}  \\
\hline \hline

\textsl{\scriptsize{Interevent intervals}}&   \scriptsize{0.7642}  &\scriptsize{0.7258}  &   \scriptsize{0.7746}  &\scriptsize{0.7907}&\scriptsize{0.7726}  \\

\textsl{\scriptsize{Separation distances}}    &\scriptsize{0.7651} &\scriptsize{0.7153} &\scriptsize{0.7871} &\scriptsize{0.7913} &\scriptsize{0.7836}\\

\textsl{\scriptsize{Depth differences}}   &\scriptsize{0.7254} &\scriptsize{0.6883} &\scriptsize{0.7772} &\scriptsize{0.7921} &\scriptsize{0.7826}\\

\textsl{\scriptsize{Magnitude differences}}&\scriptsize{0.5806}& \scriptsize{0.5521} &\scriptsize{0.5396}& \scriptsize{0.5548}&\scriptsize{0.5556}   \\

\hline

\end{tabular}
\\
\vspace{4mm} {\it Table 1.} \footnotesize{Hurst exponent values
calculated via the $CR/S$, $MR/S$ and $H_{DFA_{(1-3)}}$ methods for the time
series of intervent intervals, separation distances, depth
differences and magnitude differences.}
\end{center}

For the case of time series composed of 1115 data points,
as is the case of the four seismic time series analysed in this
work (after filtering out events that do not satisfy the Gutenberg-Richter
law), a high level of long-range correlation on the data showed up
through an overestimation of the Hurst exponents via the $CR/S$
method. In other words, the scarcity of data points in the
analysis misleads to a fictitious higher long-range correlation.
The level of long-range correlation is still not the biggest
concern as it is the accurateness of the method to establish if
there is correlation or if randomness or anti-correlation govern
the dynamics.

Although we have found an overestimation of the correlation level
in the seismic time series using the $CR/S$ method, {\it i.e.}
$H_{C}>H_{M}$, still Hurst exponents above 0.5 are obtained by
using the $MR/S$ method except for the time series of magnitude differences.
This is an important result of non-randomness for the main parameters
of a seismic event, {\it i.e.} time, and location of
seismic events, meaning they
are not arbitrary and therefore a fractal structure describes
their behaviour. In this sense, non-randomness implies the presence
of patterns behind the structure of the series and potentially a
model of future events in terms of time and location
could be possibly proposed. Nevertheless, the inclusion of a such
model is outside of the goals of this paper. Furthermore, the
level of long-range correlation is proportional to the Hurst
exponent and the trust values of {\footnotesize$H$} are obtained
via the $MR/S$ method. Values of the $H_{M}$ shown in Table 1
indicate that the persistences associated to the time series of
separation distances and interevent intervals are larger than the
ones associated to the time series of depth differences and
magnitude differences.

We employ the {\it DFA} that makes it possible to see whenever the
data are affected by different orders of polynomial trends and
with this implementation detrended Hurst exponent estimations are
feasible (see \cite{Domino2011} and  references therein). We
implement the $DFA$ until the third order to study the existence
of trends in seismic time series. This method allows removing
trends of order $n-1$ in the data and the comparison between
different $n$ orders gives an evidence of a specific polynomial
order of trends. For the case in which the Hurst exponent
calculated with {\footnotesize$DFA_{n+1}$} is smaller than the
corresponding one with {\footnotesize$DFA_n$}, higher long-range
correlations are in fact masked by the presence of trends of order
$n$.

Following the analysis via the $DFA$ until the third order, we
obtain that the analyzed time series do not present trends. We can
make this conclusion starting from the Hurst exponents that we
present in Table 1. The differences between
{\footnotesize$H_{DFA_1}$},
{\footnotesize$H_{DFA_2}$}, {\footnotesize$H_{DFA_3}$} are so small
and they are due to the precision of the algorithm and the differences
are between the interval of uncertainty of the method $\pm0.03$.

%%%%%%%%%%%%%%%%%%%%%%%%%%%%%%%%%%%%%%%%%%%%%%%%%%%%%%%%%%%%%%%%%%
%%%%%%%%%%%%%%%%%%%%%%%%%%%%%%%%%%%%%%%%%%%%%%%%%%%%%%%%%%%%%%%%%%
\section{Colombian subduction zone}
%%%%%%%%%%%%%%%%%%%%%%%%%%%%%%%%%%%%%%%%%%%%%%%%%%%%%%%%%%%%%%%%%%
%%%%%%%%%%%%%%%%%%%%%%%%%%%%%%%%%%%%%%%%%%%%%%%%%%%%%%%%%%%%%%%%%%

The highest level of seismic activity in Colombia is due to the
presence of subduction zones. The Nazca plate under the South
American plate creates the most representative origin of seismic
events in this area. Therefore, with the aim to
give more insights about the nature of the seismic fluctuations,
the area that encloses the subduction of the Nazca plate beneath
the South American plate has been selected for the study that we
present in this section, in order to account events with a common
physical nature.

We first inspect the behaviour of the Hurst exponent when the
maximum window size $M_{max}$ is varied in the calculation,
with the purpose to find the most suitable value of $M_{max}$ that
depends on the number of events $N$ considered in time.

In the second part of this section, we explore the agreement in
the time evolution between the magnitude and the Hurst exponent on
the seismic time series. For this analysis three different maximum window sizes $M_{max}$
are chosen. Meaningful results for waiting times were reported
previously for the Southern M\'exico \cite{Alvarez2012}. We extend
here these results for the case of time series of interevent
intervals associated to the Colombian subduction zone.

%%%%%%%%%%%%%%%%%%%%%%%%%%%%%%%%%%%%%%%%%%%%%%%%%%%%%%%%%%%%%%%%%%
%%%%%%%%%%%%%%%%%%%%%%%%%%%%%%%%%%%%%%%%%%%%%%%%%%%%%%%%%%%%%%%%%%
\subsection{Hurst exponent and maximum window size}

The dependence of the Hurst exponent with the maximum window size
$M_{max}$ has been widely studied for time series. However, it has
been always difficult to establish criteria that determine the
relationship between this parameter and the total number of data
points. This dependence is extremely important for the study of
the time evolution of the Hurst exponent. Varying the maximum
window size $M_{max}$ and taking a minimum window size $M_{min}$
of 5 events as a threshold for sufficient statistics, we obtain
the behaviour of the Hurst exponent for the time series of
interevent intervals in the Colombian subduction zone through the
$MR/S$ (figure 4).

\begin{figure}[H]
\begin{center}
% For example, with the graphicx package use e.g
\includegraphics[width=0.9\textwidth]{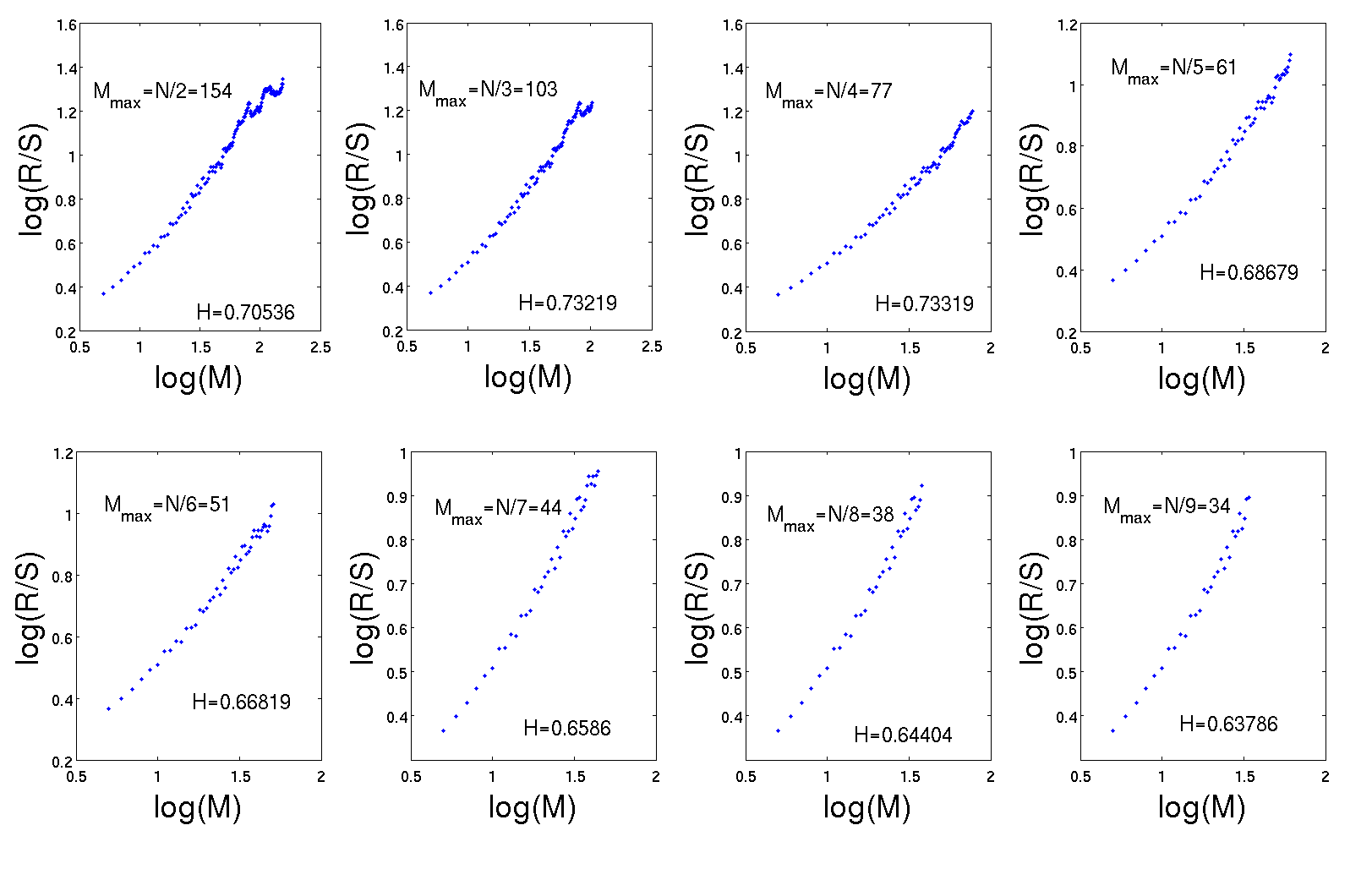}
\caption{$log(R/S)-log(M)$ plots for different $M_{max}$ for the
time series of interevent intervals in the Colombian subduction zone.
The employed method for calculations was the $MR/S$.}
\label{fig4} % Give a unique label
\end{center}
\end{figure}

In general one can choose the range of $M$ so that the
$log(R/S)-log(M)$ plot follows the linearity or that $(R/S)-(M)$
follows a power law as established in equation 12. In this sense
results from figure 4 suggest that the $M_{max}$ should be around
61, when considering the series with the total number of data
$N=309$.

As we wish to study the behaviour of the Hurst exponent as a
function of the number of events in the time $N(t)$, it will be
necessary to determine a suitable $M_{max}$ for each $N$. For that
purpose we have varied the series from 100 to 340 events and
calculated the Hurst exponent for different $M_{max}$. It is
important to clarify that as $M_{max}$ depends on the total number
of data points $N$ considered. We have calculated the Hurst
exponent for general $M_{max}=N/2, N/3, N/4, N/5, N/6, N/7, N/8,
N/9$ employing the $CR/S$ and $MR/S$ methods, as it is illustrated
in Figure 5.

When comparing results from the $CR/S$ method (Figure 5(a)) and
the $MR/S$ method (Figure 5(b)), it is possible to observe that
there is an overall constant overestimation of the Hurst exponent
calculated via the $CR/S$ method of about 0.1. The scarcity of the
seismic data points for the subduction zone in the studied years
makes the use of the $MR/S$ method an ideal one to calculate Hurst
exponents.

\begin{figure}[H]
\begin{center}
% For example, with the graphicx package use e.g
\includegraphics[width=0.45\textwidth]{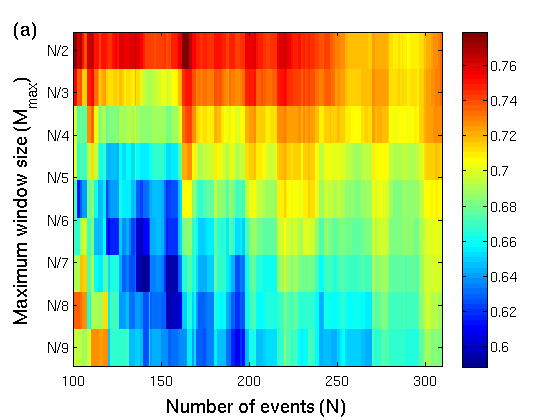}
\includegraphics[width=0.45\textwidth]{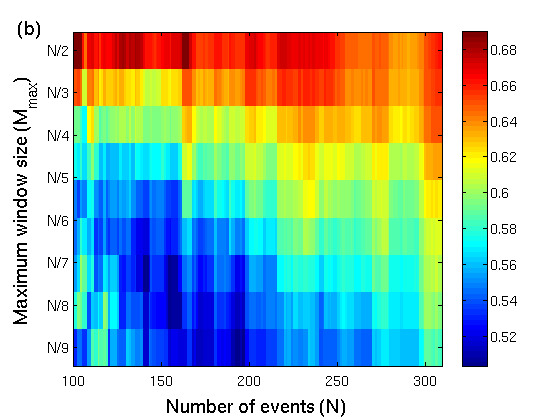}
\caption{Behaviour of the Hurst exponent under the variation of the
minimum number of windows (maximum window size) and total number
of events considered in time. {\bf(a)} Hurst exponent calculated
with the $CR/S$ method. {\bf(b)} Hurst exponent calculated using
the $MR/S$ method. }
\label{fig5} % Give a unique label
\end{center}
\end{figure}

%%%%%%%%%%%%%%%%%%%%%%%%%%%%%%%%%%%%%%%%%%%%%%%%%%%%%%%%%%%%%%%%%%
%%%%%%%%%%%%%%%%%%%%%%%%%%%%%%%%%%%%%%%%%%%%%%%%%%%%%%%%%%%%%%%%%%
\subsection{Relation between Hurst exponent and magnitude}

\begin{figure}[H]
\begin{center}
% For example, with the graphicx package use e.g
\includegraphics[width=0.45\textwidth]{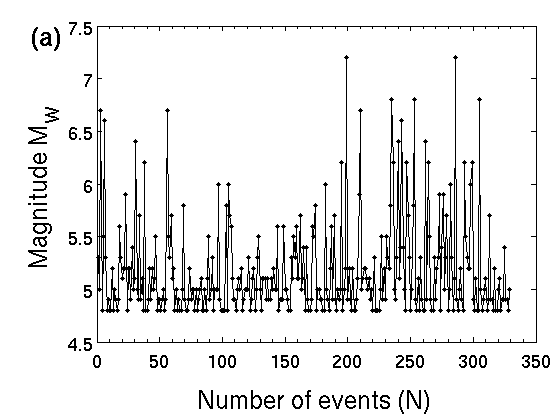}
\includegraphics[width=0.45\textwidth]{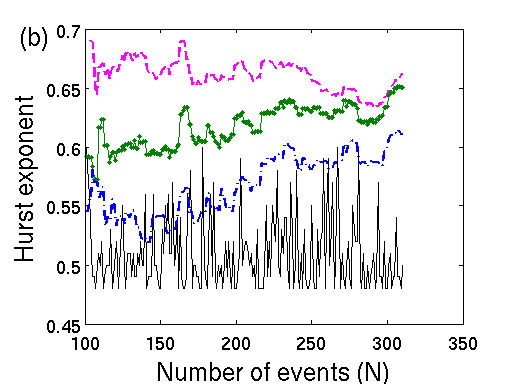}
\caption{{\bf(a)} Magnitude over the time from 0 to 350 events for
the subduction zone. {\bf(b)} Comparison between the magnitude of
seismic events divided by a factor of 10 and the Hurst exponent
calculated for the time series of interevent intervals for a range
defined from 100 to 330 events. Hurst exponents for three
different maximum window size $M_{max}$ are shown: upper
equal to $N/2$, middle equal to $N/4$ and lower equal to $N/7$.}
\label{fig6} % Give a unique label
\end{center}
\end{figure}

We analyse only those events which follow the Gutenberg-Richter
law, with magnitudes between 4.8 and 6 $M_W$. We can observe in
Figure 6(a) different areas of low magnitude that we call {\it
low-magnitude gaps}. In a previous work \cite{Alvarez2012}, it was
demonstrated that those low-magnitude gaps are closely related to
low correlation levels with values of $H$ around 0.5. In the
present work, we investigate this relation for the case of the
time series of interevent intervals. The results obtained show
that $H$ is closer to 0.5 for the {\it low-magnitude gap} between
the 100 and the 175 event (see Figure 6(b)). For the case of the
seismic events with magnitudes from 5.5 to $~$6 $M_W$, the Hurst
exponent becomes also closer to 0.5. This fact means a random
behaviour for low magnitude events.

In this work, the relationship obtained between the magnitude and
the Hurst exponent can be considered as an extension of the one
obtained in \cite{Alvarez2012}, but including several values of
the maximum window size $M_{max}$. The results that we have
obtained are shown in Figure 6(b), for three different maximum window size.
The upper curve corresponds to $M_{max}=N/2$,
where a correspondence between the magnitude and the Hurst
coefficient can not be established. The middle and the lowest
curves account $M_{max}$ values of $N/4$ and $N/7$ respectively, and these
curves show certain correlation in the behaviour of the persistence
evolution and the magnitude evolution. For both curves, there is a
gap from 114 to 174 of low Hurst exponents closer to 0.5.
Therefore, a relation between random behaviour and low seismic
activity is found for the Colombian subduction zone.

If we regard, for values smaller than $N/2$, the dependence of $H$
respect to the $M_{max}$, then the behaviour of the Hurst exponent
is reproducing more accurately the evolution on time of the
magnitude. This fact could be taken as a desirable threshold for
the maximum number of windows on the $R/S$ analysis, since
$M_{max}=N/4$ as well as $M_{max}=N/7$ reproduce more accurately the
magnitude dynamics than $M_{max}=N/2$.

\section{Conclusions}

The seismic time series of interevent intervals, separation
distances and depth differences are found to be persistent via the
$CR/S$ method. Nevertheless, using the $MR/S$ method, it is
determined that an overestimation of the Hurst exponent $H$
affects all the series due to the scarcity of data points (around
2000). Despite the overestimation mentioned above, all the time
series under consideration except for the time series of magnitude
differences present persistence with $0.5<H_G<1$, when corrections
via the $MR/S$ method are taken into account.

Trends were not detected via the comparison between different
hierarchies of the $DFA$ until third order for the four time
series studied. The differences between {\footnotesize$H_{DFA_1}$},
{\footnotesize$H_{DFA_2}$}, {\footnotesize$H_{DFA_3}$} are attributed
to the precision of the algorithm.

For the time series of interevent intervals, the analysis of the
Hurst exponent as a function of the time and the minimum number of
windows has shown that $M_{max}<N/2$ is ideal for calculations,
since it is describing better the dynamics of seismic events.
Furthermore, we have found an agreement in the time evolution
between low magnitude events and the value of Hurst exponent
closer to 0.5 for the gap between 114 and 174 events.


\begin{thebibliography}{999}
\bibitem{Stanley1971}
H. E. Stanley, {\it Introduction to phase transitions and critical
phenomena}, Oxford University Press, London, 1971.

\bibitem{Mantegna2000}
R. N. Mantegna, and H. E. Stanley, {\it An introduction to
econophysics: Correlations and complexity in finance}, Cambridge
University Press, Cambridge, 2000.

\bibitem{Grau2000}
P. Grau-Carles, {\it Empirical evidence of long-range correlations
in stock returns}, Physica A 287 (2000) 396-404 .

\bibitem{Varotsos2002}
P. A. Varotsos, N. V. Sarlis, and E. S. Skordas, {\it Long-range
correlations in the electric signals that precede rupture}, Phys.
Rev. E 66 (2002) 011902.

\bibitem{Forrest1979}
S. R. Forrest, and T. A. Witten Jr., {\it Long-range correlations
in smoke-particle aggregates}, J. Phys. A: Math. Gen. 12 (1979)
L109-L117.

\bibitem{Peng1992}
C. -K. Peng, S. V. Buldyrev, A. L. Goldberger, S. Havlin, F.
Sciortino, M. Simons, and H. E. Stanley, {\it Long-range
correlations in nucleotide sequences}, Nature 356 (1992) 168-170.

\bibitem{Hirabayashi1992}
T. Hirabayashi, K. Ito, and T. Yoshii, {\it Multifractal analysis
of earthquakes}, Pure Appl. Geophys. 138 (1992) 591-610.

\bibitem{Peng1994}
C. -K. Peng, S. V. Buldyrev, S. Havlin, M. Simons, H. E. Stanley,
and A. L. Goldberger, {\it Mosaic organization of DNA
nucleotides}, Phys. Rev. E 49 (1994) 1685.

\bibitem{Ebeling1994}
W. Ebeling, and T. Poschel, {\it Entropy and long range
correlations in literary English}, Europhys. Lett. 26 (1994)
241-246.

\bibitem{Hausdorff1995}
J. F. Hausdorff, C. -K. Peng, Z. Ladin, J. Y. Wei, and A. L.
Goldberger, {\it Is walking a random walk? Evidence for long-range
correlations in stride interval of human gait}, J. of Appl.
Physiol. 78 (1995) 349-358.

\bibitem{Peng1995}
C. -K. Peng, S. Havlin, H. E. Stanley, and A. L. Goldberger, {\it
Quantification of scaling exponents and crossover phenomena in
nonstationary heartbeat time series}, Chaos 5 (1995) 82-87.

\bibitem{Liu1997}
Y. Liu, P. Cizeau, M. Meyer, C. -K. Peng, and H. E. Stanley, {\it
Correlations in economic time series}, Physica A 245 (1997)
437-440 .

\bibitem{Segre1997}
P. N. Segre, E. Herbolzheimer, and P. M. Chaikin, {\it Long-range
correlations in sedimentation}, Phys. Rev. Lett 79 (1997) 2574 .

\bibitem{Koscielny1998}
E. Koscielny-Bunde, A. Bunde, S. Havlin, H. Eduardo Roman, Y.
Goldreich, and H. J. Schellnhuber, {\it Indication of a universal
persistence law governing athmospheric variability}, Phys. Rev.
Lett. 81 (1998) 729-732.

\bibitem{Blesic1999}
S. Blesic, S. Milosevic, D. Stratimirovic, and M. Ljubisavljevic,
{\it Detrended fluctuation analysis of time series of a firing
fusimotor neuron}, Physica A 268 (1999) 275-282.

\bibitem{Ivanova1999}
K. Ivanova, and M. Ausloos, {\it Application of the detrended
fluctuation analysis (DFA) method for describing cloud breaking},
Physica A 274 (1999) 349-354.

\bibitem{Telesca2004-1}
L. Telesca, V. Lapenna, and M. Macchiato, {\it Mono- and
multi-fractal investigation of scaling properties in temporal
patterns of seismic sequences}, Chaos Sol. Frac. 19 (2004) 1-15.

\bibitem{Telesca2004-2}
L. Telesca, and M. Macchiato, {\it Time-scaling properties of the
Umbria-Marche 1997-1998 seismic crisis, investigated by the
detrended fluctuation analysis of intervent time series}, Chaos
Sol. Frac. 19 (2004) 377-385.

\bibitem{Gavin2009}
S. Gavin, L. McLerran, and G. Moschelli, {\it Longe range
correlations and the soft ridge in relativistic nuclear
collisions}, Phys. Rev. C 79 (2009) 051902(R).


\bibitem{Ashkenazy}
Y. Ashkenazy, S. Havlin, P. Ch. Ivanov, C.K.  Peng,
V. Schulte-Frohlinde,  and H. E. Stanley, {\it Magnitude
and sign scaling in power-law correlated time series}, Physica
A 323 (2003) 19-41.


\bibitem{Telesca2008}
L. Telesca, and M. Lovallo, {\it Investigating non-uniform
scaling behaviour in temporal fluctuations of seismicity}, Nat. Hazards Earth Syst. Sci. 8 (2008) 973-976.



\bibitem{Michas}
G. Michas , F. Vallianatos , and P. Sammonds, {\it Non-extensivity and long-range correlations in the
earthquake activity at the West Corinth rift (Greece)}, Nonlin. Processes Geophys. 20 (2013) 713-724.


\bibitem{Feder1988}
J. Feder, {\it Fractals}, Plenum Press, New York, 1988.

\bibitem{Chen2008}
C. Chen, Y. Lee, and Y. Chang, {\it A relationship between Hurst
exponents of slip and waiting time data of earthquakes}, Physica A
387 (2008) 4643-4648.

\bibitem{Mandelbrot1968}
B. B. Mandelbrot, and J. R. Wallis, {\it Noah, Joseph and the
operational hydrology}, Water Resour. Res. 4 (1968) 909-918.

\bibitem{Hurst1951}
H. E. Hurst, {\it Long-term storage capacity of reservoirs},
Trans. Am. Soc. Civil Eng. 116 (1951) 770-808.

\bibitem{Mandelbrot1969-1}
B. B. Mandelbrot, and J. R. Wallis, {\it Some long-run properties
of geophysical records}, Water Resour. Res. 5 (1969) 321-340.

\bibitem{Mandelbrot1969-2}
B. B. Mandelbrot, and J. R. Wallis, {\it Robustness of the
rescaled range R/S in the measurement of noncyclic long-run
statistical dependence}, Water Resour. Res. 5 (1969) 967-988.

\bibitem{Sanchez2008}
M. A. S{\'a}nchez Granero, J. E. Trinidad Segovia, and J.
Garc{\'i}a P{\'e}rez, {\it Some comments on Hurst exponent and the
long memory processes on capital markets}, Physica A 387 (2008)
5543-5551.

\bibitem{Trinidad2012}
J. E. Trinidad Segovia, M. Fern{\'a}ndez-Mart{\'i}nez, and M. A.
S{\'a}nchez Granero, {\it A note on geometric method-based
procedures to calculate the Hurst exponent}, Physica A 391 (2012)
2209-2214.

\bibitem{Lo}
W. Lo, {\it Long-Term Memory in Stock Market Proces}, Econometrica 59 (1991)
98-109.

\bibitem{Kantelhardt2001}
J. W. Kantelhardt, E. Koscielny-Bunde, H. H. Rego, S. Havlin, and
A. Bunde, {\it Detecting long-range correlations with detrended
fluctuation analysis}, Physica A 295 (2001) 441-454.

\bibitem{Hu2001}
K. Hu, P. Ch. Ivanov, Z. Chen, P. Carpena, and H. E. Stanley, {\it
Effect of trends on detrended fluctuation analysis}, Phys. Rev. E
64 (2001) 011114.

\bibitem{Eichner2003}
J. F. Eichner, E. Koscielny-Bunde, A. Bunde, S. Havlin, and H. -J.
Schellnhuber, {\it Power-law persistence and trends in the
atmosphere: A detailed study of long temperature records}, Phys.
Rev. E 68 (2003) 046133.

\bibitem{Kantelhardt2002}
J. W. Kantelhardt, S. A. Zschiegner, E. Koscielny-Bunde, S.
Havlin, A. Bunde, and H. E. Stanley, {\it Multifractal detrended
fluctuation analysis of nonstationary time series}, Physica A 316
(2002) 87-114.

\bibitem{Telesca2007}
L. Telesca, M. Lovallo, V. Lapenna, and M. Macchiato, {\it
Long-range correlations in two-dimensional spatio-temporal sismic
fluctuations}, Physica A 377 (2007) 279-284.

\bibitem{Alvarez2012}
J. Alvarez-Ramirez, J. C. Echeverria, A. Ortiz-Cruz, and E.
Hernandez, {\it Temporal and spatial variations of seismicity
scaling behavior in Southern Mexico}, J. of Geodyn. 54 (2012)
1-12.

\bibitem{Hirata1989}
T. Hirata, {\it A correlation between the b value and the fractal
dimension of earthquakes}, J. of Geophys. Res. 94 (1989)
7507-7514.

\bibitem{Guo1995}
Z. Guo, and Y. Ogata, {\it Correlation between characteristic
parameters of aftershock distributions in time, space and
magnitude}, Geophys. Res. Let. 22 (1995) 993-996.

\bibitem{Guo1997}
Z. Guo, and Y. Ogata, {\it Statistical relations between the
parameters of aftershock distributions in time, space and
magnitude}, J. of Geophys. Res. 102 (1997) 2857-2873.

\bibitem{Varotsos2012}
P. A. Varotsos, N. V. Sarlis, and E. S. Skordas, {\it
Scale-specific order parameter fluctuations of seismicity before
mainshocks: Natural time and detendred fluctuation analysis}, EPL
99 (2012) 59001.

\bibitem{usgs}
U.S. Geological Survey, {Earthquake data base}, USGS, accessed
[Oct. 31, 2011] at URL
[http://earthquake.usgs.gov/earthquakes/search/].

\bibitem{Jimenez2011}
A. Jimenez, {\it Comparison of the Hurst and DEA exponents between
the catalogue and its clusters: The California case}, Physica A
390 (2011) 2146-2154.

\bibitem{Burton2006}
Y. Xu, and P. W. Burton, {\it Time varying seismicity in Greece:
Hurst's analysis and Monte Carlo simulation applied to a new
earthquake catalogue for Greece}, Tectonophysics 423 (2006)
125-136.

\bibitem{Domino2011}
K. Domino, {\it The use of the Hurst exponent to predict changes
in trends on the Warsaw Stock Exchange}, Physica A 390 (2011)
98-109.

\end{thebibliography}
\end{document}